Michał BRZEZIŃSKI-SPICZAK[*], Krzysztof DOBOSZ[*],
Mateusz LIS[*], Łukasz PINTAL[*]

# MUSIC FILES SEARCH SYSTEM

This paper introduces a project of advanced system of music retrieval from the Internet. The system uses combination of text search (by author, title and other information about the music file included in *id3 tag* description or similar for other file types) with more intuitive and novel method of melody search using *query by humming*. The patterns for storing text and melody information as well as improved clustering algorithm for the pattern space were proposed. The search engine is planned to optimise the query due to the data input by user, thanks to the structure of text and melody index database. The system is planned to be a plug-in for popular digital music players or an independent player. An advanced system of recommendation based on information gathered from user's profile and search history is an integral part of the system. The recommendation mechanism uses *scrobbling* methods and is responsible for making suggestions of songs unknown to the user but similar to his preferred music styles and positioning search results.

## 1. INTRODUCTION

Everyday the number of music files available on the Internet increases. The appearance of community portals and file exchange systems has got a major impact on already vast number of that kind of data. Thus, it is a necessity to introduce advanced music search techniques. Although there are many papers dealing with the problem of creating data structure representing music information that supports accurate and effective searching (eg. [5], [6], [7]), it is hard to find a complete approach to design a music search and storage engine. Of course, there are commercial systems that allow to find music files on the Internet, nevertheless most of them is still based on text indexing

---
[*] Institute of Information Science and Engineering, Wrocław University of Technology, Janiszewskiego 11/17 50-370 Wrocław, Poland

using tag description created by users. It seems to be obvious, trying to find a song by its title, author or album on which it appeared. However, this method may be not sufficient in some specific cases when the user simply does not have such detailed information. On the other hand, there are many other kinds of information that describe a piece of music, such as snippets of lyrics, context in which the song appeared on the web or even melody fragments. Probably the most intuitive searching method, from the users point of view, is *query by humming* which allows to find a piece of music without knowing any text information about it ([5], [7]). The main goal of this paper is to introduce a complete music search system based on both text and melody indexing methods. The text information includes context of tune appearance. In this case an associating system, based on external text searching engine (such as Google™), was introduced. In addition, a tune recommendation mechanism based on the *scrobbling* method was developed and described, to allow the system to sort the search results properly and accurately for individual user.

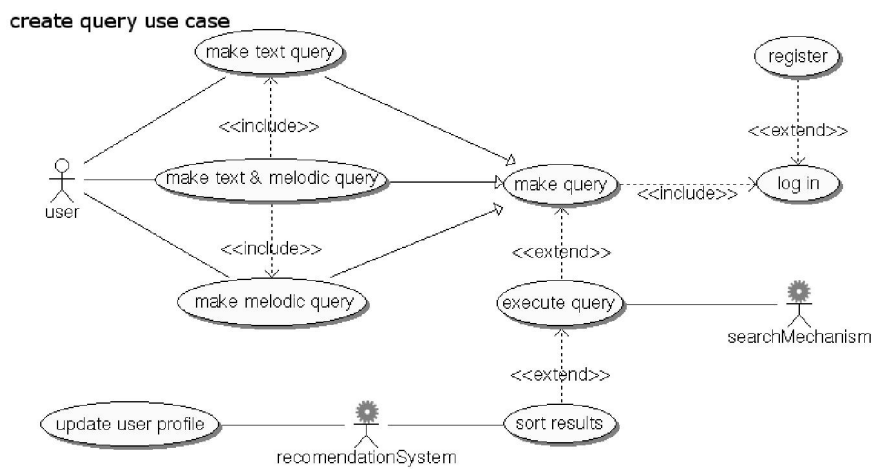

Fig. 1. UML diagram of the main use case of the system.

## 2. INDEXING METHODS

### 2.1. TEXT-INDEXING

Majority of the music search systems is based on text *tags* description. In most cases such method is sufficient to locate the file and when it is not, it may be very helpful anyway. The most commonly used queries consist of the title, authors name, album title, music genre or release year. This type of data may be retrieved from *id3 tags* of the *mp3* files or similar descriptions of other file formats. Moreover, lyrics snippets are also often used as a search queries. Other types of queries such as nationality of the composer, lyrics language, record company could be used in some paricular cases, i.e. when a user is interested in Slovakian folk music. All of the mentioned data needs to be stored and indexed.

Another element of the index needs to cover the problem of the so-called *free associations*, which means expressions characteristic for the specific tune but impossible to categorise as any of the mentioned before. An association might be name of a product, which was advertised using the tune or 'Live Aid' if the song was performed on a big charity event. Such associations are to be extracted by the ***associating mechanism***.

Main principle of the mechanism is based on the assumption that every association is represented by closeness of expresions on the websites. Furthermore, it is assumed that accurate external text search engine is available. A metric function *d(expression1,expression2)* may be proposed to describe the distance between two expresions on the website using number of words between them. First step of associating is creating queries for the text search engine consisting of all known data. Then, if a particular word appears in more than 1 per cent of the result pages, it will be associated with the considered tune. For every association matching function may be calculated:

$$\delta(tune, word) = \alpha \frac{N}{I} + \beta \frac{\sum_{i=1}^{I} d(word, pattern(i))}{I} + \gamma k \qquad (1)$$

Where:
*N* – number of all results from the text search,
*I* – number of results containing considered association,
*k* – number of the first result in which the association appeared,
 $\alpha$, $\beta$, $\gamma$ – arbitrarily set parameters

All associations extracted from the web with $\delta$ function values are stored and indexed in the system. The function $\delta$ is used to obtain the order of search results.

## 2.2. MELODIC INDEXING

The analysis of melody characteristics shows that information about the height, duration and the ring time of a following notes is enough for successful recognition of a piece of music [5]. This fact may be fully confirmed by intuition – it is easy to recognise a melody although it is played on a different instrument or when its tempo has been changed. A piece of music can be described using series of triplets, each associated with a single note.

Of course there is a problem of poliphony in pieces of music. One tune is considered as a series of sequences of sounds. The simplest way of dealing with that problem is considering every melody line separately, although this approach seems to be highly uneffective. Proposed method is to flatteren the tune to one leading melody line. Flattering consists of extracting the highest and loudest sounds at the time. Moreover, it is necessary to remove drums line because of its inability to contain melodies and dominating character in the tune.

Because of such characteristic three major methods of melody representation can be proposed:
- *PIT – the pitch interval with the previous note*
- *IOI – the interonset interval with the following note*
- *BTH – both PIT and IOI*

Each method will be presented on the example of a fragment of *Ode to Joy* by Ludwig von Beethoven.

Fig. 2. Fragment of *Ode to Joy* by Ludwig von Beethoven.

The *PIT* description is based on height differences between following notes which corresponds with frequencies of sounds that may be extracted directly from the music file. In its basic form, *PIT* is a formated string containing pitch differences between notes measured in half-tones. Such approach, however, is highly susceptible to errors which may often occure during singing or humming. This leads to another, quantified form of *PIT* which is recommended for practical purposes [7]. In this form the only information used is if the following note is higher or lower than the previous one.
The fragment of melody presented above transcripted to basic *PIT* is the following:
```
* 0 2 -4 2 2 1 -1 -4 2 2 1 -1 -2 -2 2 -7
```

```
     9  0  1  2  0 -2 -1 -2 -2  0  2  2 -2 -2  0
```
The same fragment transcripted to quantified form:
```
     *  0  +  -  +  +  +  -  -  +  +  +  -  -  -  +  -
     +  0  +  +  0  -  -  -  -  0  +  +  -  -  0
```

The interonset interval with the following note form is similar to *PIT*, although it is based on ring times and duration instead of note heights. It is important to notice, that duration is less important than a time distance between the onsets of the notes.

It is not easy to define a proper unit for such transcription form. There is no elemental unit similar to half-tone in *PIT* for time distance analysis. A proposed unit, which is an average of *k* shortest time intervals, allows to avoid humming errors. Parameter *k* is arbitrarily set. What is more, there is a possibility of introducing a quantified notation of *IOI* which is similar to quantified *PIT*.

Transcripted *IOI* form of given melody ($k=4$):
```
     2  2  2  2  1  1  2  2  2  1  1  2  2  2  2  4
     2  2  2  2  2  2  2  2  2  2  2  2  2  1  4
```
Similarly, the quantified *IOI* form is:
```
     *  0  0  0  0  -  0  +  0  0  -  0  +  0  0  0  +
     -  0  0  0  0  0  0  0  0  0  0  0  0  -  +
```

*BTH* notation is a join of *PIT* and *IOI* and is based on quantified form of those two notations. It was first introduced in [5] and proved to be useful. *BTH* consists of tuples representing *PIT* and *IOI* notation of each note.

Example of *BTH* transcription of *Ode to Joy*:

(\*,\*) (0,0) (+,0) (-,0) (+,-) (+,0) (-,+) (-,0) (+,0) (+,-) (+,0) (-,+) (-,0) (-,0) (+,0) (-,+)
(+,-) (0,0) (+,0) (+,0) (0,0) (-,0) (-,0) (-,0) (-,0) (0,0) (+,0) (+,0) (-,0) (-,-) (0,+)

Despite simplified representations of music data, transcriptions of whole music files are still too large to be effective. The solution of the problem is storing so-called *patterns*, which are choruses and other characteristic parts of music transcription (usually the most common fragments of melody), as indexes of full transcriptions in database.

Another worth mentioning characteristic of the tune is its *personality*, which stands for variability of whole pattern. In probabilistic meaning it is standard deviation of note heights and onset interval lenghts.

## 3. DATA STORAGE STRUCTURE

### 3.1. METRIC SPACE OF MELODIES

Simplified notation of music data allows to introduce measures of similarity. Because of the fact that *PIT*, *IOI* or *BTH* are text strings, measure may be based on edit distance described in [2]. The main idea of this measures is to assign proper factors to elementary editing operations such as adding, deleting or replacing characters and computing minimum number of those operations needed to convert one string into part of another, multiplied by corresponding factors. The calculated value is used as a measure of similarity. To sum up, for every notation of pattern there is a different metric space based on the proposed measure.

### 3.2. PARTITIONS IN MELODIC SPACE

Due to the vast amount of music data stored in database, it is required to clusterise metric space of melody patterns and create indexes of clusters to achieve satisfying search performance. *K-means*, described in [3], may be used to deal with the problem, nevertheless it requires to specify the number of clusters *a priori*, which in considered case is hard to obtain. Thus, another clustering algorithm shall be suggested.

Suppose that large ammount of points in metric space is considered. First step is to find the most distant point and create a cluster of point placed not further than arbitrarily set parameter $d_0$. Then, again most distant point is obtained in the metric space omiting already assigned part. This step repeats until the whole space is covered. After that, as in *K-means* algorithm, new cluster means are calculated. In their neighbourhood new clusters are created of points not further than $d$, where $d$ value is defined as maximum of $d_0$ and the distance between current cluster mean and the most distant point chosen in the first step. If a mean point of any cluster is inside another cluster, the clusters are joined and the $d$ value is recalculated. This repeats until mean points stabilise.

Repartitioning, because of its complexity, is planed to be made once a fixed period of time, while new tunes are added using *nearest neighbour* algorithm.

### 3.3. DATABASE STRUCTURE

To sum up, index of the tune consist of information about:
- tune – title, lyrics, genre
- artist – author of music, lyrics, performer
- release – album, single, or concert bootleg
- performance – one song may be performed by different artists during various events
- release company
- free associations
- melody indexes – including *PIT*, *IOI* and *BTH*.

This leads to text and melody index database structure presented in fig. 3.

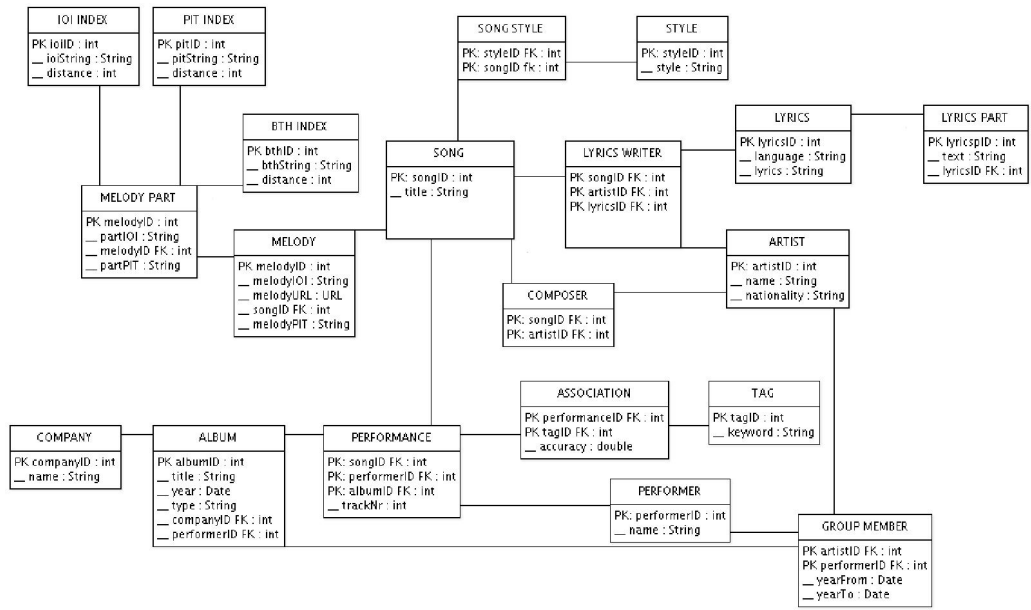

Fig. 3. Structure of database.

## 4. SEARCH ALGORITHMS

### 4.1. SEARCHING USING TEXT TAGS

Text indexes in database are represented by lexical trees. All data connected with the tune (including free associations) is stored in this structure. Expressions extracted from text query are binary compared to every tree. Order of searched trees is determined by additional informations from the query, i.e. "[ALBUM]Californication" induces searching firstly in the album tree. More information about lexical trees and text search algorithms can be found in [1].

In case of more complex queries consisting of more than one expression, search results are subjective to binary logical constraints, where default operator is AND.

### 4.2. SEARCH IN MELODIC SPACE METHODS

First of all, *PIT*, *IOI* and *BTH* patterns extracted from users query are compared to mean points of every cluster in melody space. Clusters with means not further than arbitrarily set value $d_1$ are united as a search list. All selected points are sorted by their distance to the query pattern. The sorted list is returned as a search result. The formal notation of the algorithm is presented below:

$P$ – set of clusters in melody space
$p_i \in P$ – single cluster
$\omega(p_i)$ – mean point of the cluster
$\omega$ – search pattern extracted from users query
$d(\omega_1, \omega_2)$ – distance between patterns $\omega_1$ and $\omega_2$

$$Z(P, \omega_0, d_1) = \{\omega : \omega \in p_i, p_i \in P, d(\omega(p_i), \omega_0) \leq d_1\}$$

1. Create empty result list $L$
2. Extract the pattern $\omega_i$ from the query
3. Obtain $Z(P, \omega_i, d_1)$
4. $\forall \omega \in Z(P, \omega_i, d_1)$ add $(\omega, d(\omega, \omega_i))$ to $L$
5. If the search query is not empty go to 2.
6. Sort $L$ by $d(\omega, \omega_i)$
7. Return $L$

### 4.3. FINAL SEARCH ENGINE

The best results will be achieved by using both search methods. In that case the search process proceeds as follows: at first the melody space is constrained to results of text search and then melody search is applied. That king of join of search methods is much more efficient than using them separately or paralelly.

Another important problem is the formal description of the query language for usage of the joined search system. The proposed language is based on binary logic, expressions will have the following form:

$$Z = Op_1[p_1]W_1 Op_2[p_2]W_2 ... Op_n[p_n]W_n[\$M_1][\$M_2]...[\$M_m] \qquad (2)$$

Where:
$Z$ – correct search query,
$W_i$ – text search expression consist of alphanumeric characters,
$Op_i$ – logical operator ('and' - default, '!', 'or'),
$p_i$ – search parameter ([TITLE], [ARTIST], [LYRICS], [ALBUM] or empty),
$M_i$ – music query converted to *PIT*, *IOI* and *BTH* notation.

It is woth mentioning that to assure proper functioning of the search engine, whole query has to be transformed to disjunctive normal form with all negations moved to the end of query.

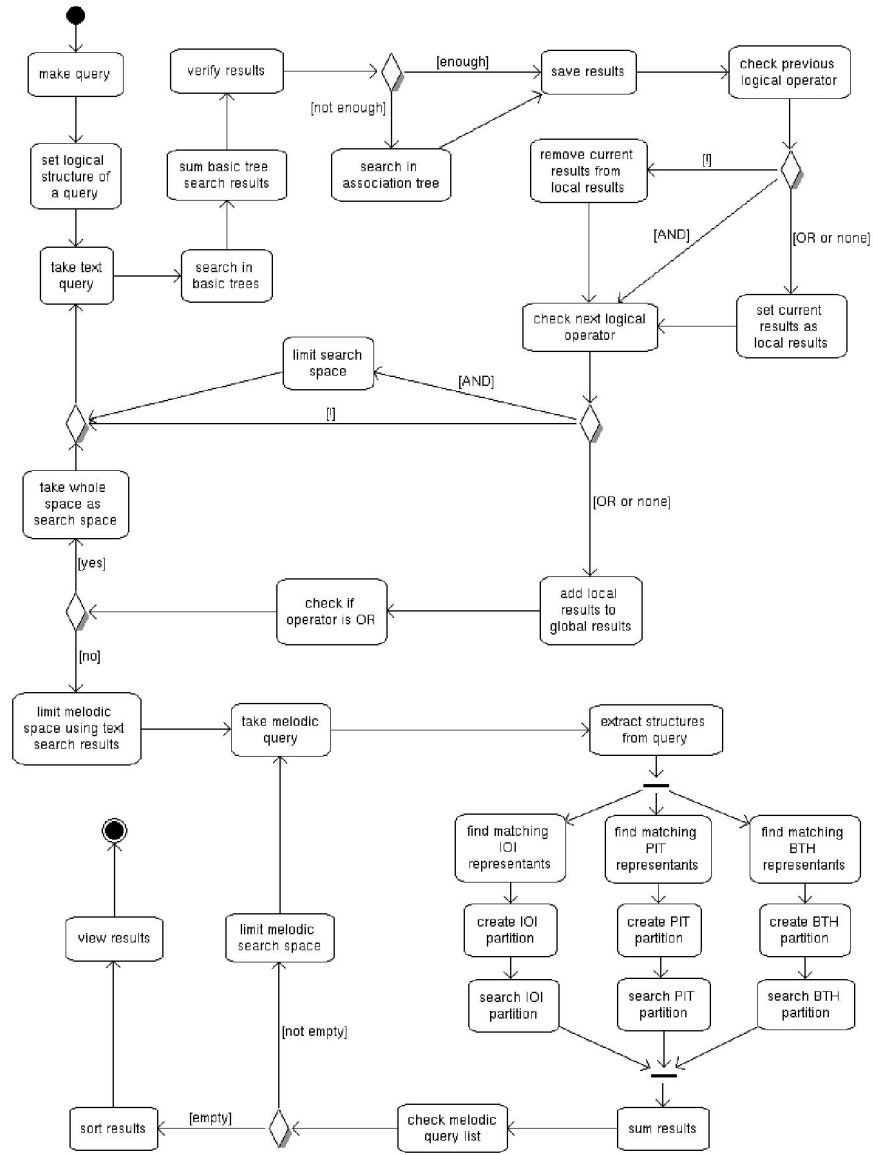

Fig. 4. Search activity diagram.

## 5. SEARCH RESULTS PRESENTATION

### 5.1. RECOMMENDATION SYSTEM

Recommendation systems known from other kinds of web services, such as Internet shops, cannot be applied to music environment because that kind of approach would require constant confirmation of search results from user. The proposed method is based on streaming broadcast stations (which is similar to thematic radiostations). This approach uses *scrobbling* technique. It means transfering elementary information about the tune currently listened by the user to the system and adding it to users profile. The most important technical requirement is creating a plug-in for users favourite music player which would be able to transfer that kind of data. Analysis of listened tunes completed by history of searches and static data like users age or sex, allows to create wide user interests profile. Users may be assigned to groups by their profile information, the system might suggest them new tunes listened by other group members helping to expand their music interest area.

### 5.2. RESULTS ORDER

Information stored in users profile may also be helpful in obtaining proper search results order. Position of a tune on results list not only depends on accuracy of query but also on the tunes popularity, declared music preferences and other group members searches (or even scrobbled tunes). Thus, the final result list shall be sorted by a factor defined as follows.

$$relevancy = \alpha \frac{1}{d(query, index)} + \beta gen + \gamma \sum listened + \delta pop \qquad (3)$$

Where:
*d(query, index)* – minimal distance in melody space to music query expression
*gen* – indicator of fulfilling users preferences
*listened* – indicator of listening to the tune by other group members extracted from *scrobbling* system
*pop* – popularity rank defined as a number of listenings of the tune by all users, also retreived using *scrobbling*
$\alpha, \beta, \gamma, \delta$ – arbitrarily set parameters

## REFERENCES

[1] CORMEN T., LEISERSON C., RIVEST R., STEIN C., *Wprowadzenie do algorytmów*, WNT, Warszawa 2004

[2] CORMODE G., MUTHUKRISHNAN S., ***The string edit distance matching problem with moves***, ACM, New York 2004.

[3] HARTIGAN A., WONG M. A. *A K-Means Clustering Algorithm*, Applied Statistics 28 (1): 1979, 100–108.

[4] MACQUEEN J.B. *Some Methods for classification and Analysis of Multivariate Observations*, Proceedings of 5-th Berkeley Symposium on Mathematical Statistics and Probability, Berkeley, University of California Press, 1967, 1:281-297.

[5] NEVE G., ORIO N., *Indexing and retreival of music documents through pattern analysis and data dusion techiques*, University of Padova, Departmend of Information Engineering, Padova 2004.

[6] PIENIMAKI A., *Indexing Music Databases using Automatic Extraction of Frequent Phrases*, University of Helesinki, Department of computer Science, Helsinki 2002.

[7] WÓJCIK J., *Methods od Forming and Ranking Rhytmic Hypotheses in Musical Pieces*. Doctoral Thesis, Gdańsk University of Technology, Faculty of Electronics, Telecomunications & Informatics, Gdańsk 2006.